# Optical properties of periodic, quasi-periodic, and disordered one-dimensional photonic structures

Michele Bellingeri[1], Alessandro Chiasera[2], Ilka Kriegel[3], Francesco Scotognella[4,5]

[1] Dipartimento di Fisica e Scienze della Terra, Viale Usberti 7/A, 43100 Parma, Italy

[2] IFN - CNR CSMFO Lab. & FBK CMM, via alla Cascata 56/C Povo, 38123 Trento, Italy

[3] Department of Nanochemistry, Istituto Italiano di Tecnologia (IIT), via Morego, 30, 16163 Genova, Italy

[4] Dipartimento di Fisica, Istituto di Fotonica e Nanotecnologie CNR, Politecnico di Milano, Piazza Leonardo da Vinci 32, 20133 Milano, Italy

[5] Center for Nano Science and Technology@PoliMi, Istituto Italiano di Tecnologia, Via Giovanni Pascoli, 70/3, 20133, Milan, Italy

*Corresponding author: francesco.scotognella@polimi.it

**Abstract**

Photonic structures are building blocks for many optical applications in which light manipulation is required spanning optical filtering, lasing, light emitting diodes, sensing and photovoltaics. The fabrication of one-dimensional photonic structures is achievable with a variety of different techniques, such as spin coating, sputtering, evaporation, pulse laser deposition, or extrusion. Such different techniques enable facile integration of the photonic structure with many types of devices. Photonic crystals are characterized by a spatial modulation of the dielectric constant on the length scale of the wavelength of light giving rise to energy ranges where light cannot propagate through the crystal – the photonic band gap. While mostly photonic crystals are referred to as periodic arrangements, in this review we aim to highlight as well how aperiodicity and disorder affects light modulation. In this review article, we introduce the concepts of periodicity, quasi-periodicity, and disorder in photonic crystals, focussing on the one-dimensional case. We discuss in detail the physical peculiarities, the fabrication techniques, and the applications of periodic, quasi-periodic, and disorder photonic structures, highlighting how the degree of crystallinity matters in the manipulation of light. We report different types of disorder in 1D photonic structures and we discuss their properties in terms of light transmission. We discuss the relationship between the average total transmission, in a range of wavelengths around the photonic band gap of the corresponding photonic crystal, and the homogeneity of the photonic structures, quantified by the Shannon index. Then we discuss the light transmission in structures in which the high refractive index layers are aggregated in clusters following a power law distribution. Finally, in the case of structures in which the high refractive index layers are aggregated in clusters with a truncated uniform distribution, we discuss: i) how different refractive index contrast tailors the total light transmission; ii) how the total light transmission is affected by the introduction of defects made with a third material.

# 1. Introduction

In the last years many researchers have reported various works on the light propagation through photonic structures. Photonic structures can be grouped in three sets, depending on their crystallographic properties: i) a periodic spatial modulation of the dielectric constant gives rise to a photonic crystal [1–9]; a modulation of the dielectric constant that follows a deterministic generation rule results in a photonic quasicrystal [10–19]; a random modulation of the dielectric constant gives rise to a disordered photonic structure [20,21]. Myriads of possible applications arise for the three different structures, such , as photonic crystal lasers [22,23] and quasicrystal lasers [24,25], optical fibers [26,27], and sensors [28–30], when concerning photonic crystals and quasicrystals. Instead, focussing on disordered photonic structures, a variety of interesting features have been discovered in different fields, as random lasing [31–37], diffuse optical imaging [38], and light harvesting for solar devices [39–42]. Many physical effects have been observed in one-dimensional disordered photonic structures, as the Anderson localization of light [43,44], the optical Bloch oscillations and necklace states [45–48], and an interesting oscillation of the average light transmission as a function of the sample length [49]. In the next three subparagraphs we will briefly introduce the topological aspects of photonic crystals, photonic quasicrystals, and disordered photonic structures:

## 1.1. 1D, 2D, and 3D photonic crystals

The dielectric function (and, consequently, the refractive index) can be periodically modulated in one-, two-, and three dimensions (Figure 1).

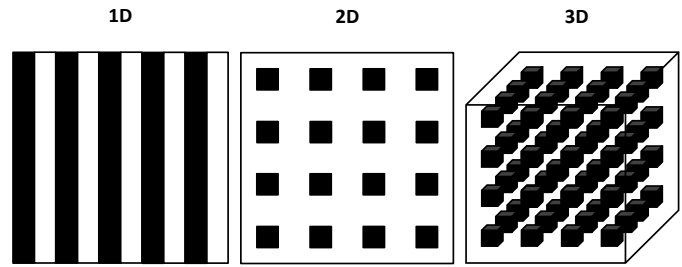

**Figure 1.** 1D (left), 2D square (center), and 3D simple cubic (right) photonic crystals.

The 1D photonic crystal is an alternated sequence of two materials with different refractive indexes. Instead, 2D and 3D crystals can be designed taking into account different types of symmetries [50]. In Figure 1, for example, we depict a 2D square lattice and a 3D simple cubic lattice.

## 1.2. 1D, 2D, and 3D photonic quasicrystals

A quasicrystal does not show a translational symmetry as periodic photonic crystals do, but is generated by a substitution rule that is based on two building blocks, resulting in a long-range order [17].

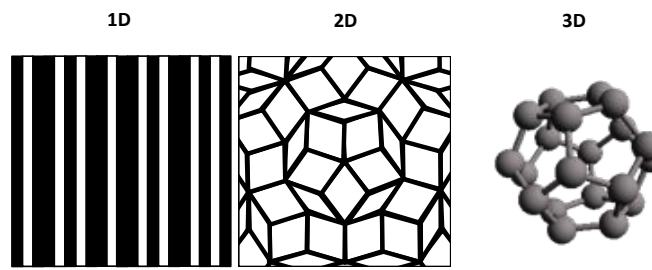

**Figure 2.** (left) 1D photonic quasicrystal following the Fibonacci sequence ABAABABAABAABABAABABA. (center) 2D Penrose photonic quasicrystal. (right) 3D icosahedral quasicrystal.

There are many different types of photonic quasicrystals [19]. For example, the 1D photonic quasicrystal in Figure 2 is generated by following the Fibonacci sequence [51–53], while the 2D photonic quasicrystal in the centre of the Figure is a Penrose structure [24,54,55]. Finally, the 3D photonic quasicrystal depicted in Figure 2 is an icosahedral quasicrystalline structure [56,57].

*1.3. 1D, 2D, and 3D disordered photonic structures*
The disordered (or random) photonic structures are the optical analogue of electronic amorphous materials [58]. Neither a short-range nor a long-range order is expected to be observed in disordered photonic structures.

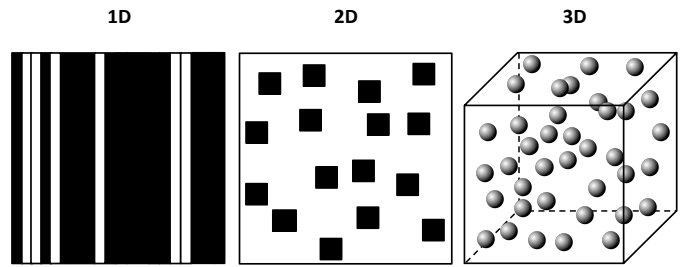

**Figure 3.** 1D (left), 2D (center), and 3D (right) random photonic structures.

In Figure 3 we show a 1D photonic structure obtained by a random sequence of two materials (depicted in black and white in the figure), a 2D structure where squares of a high refractive index material (black in the figure) are randomly dispersed in the low refractive index matrix, and a 3D structure in which high refractive index spheres are randomly dispersed in a matrix. Of course, the 2D and 3D structures can be made with different shapes and sizes of objects dispersed in the matrix [20].

In this review article we focus on 1D periodic, quasiperiodic, and disordered photonic structures. We describe some theoretical methods to study the structures, also discussing the refractive indexes of the most used dielectric materials. We discuss the optical response, the fabrication methods, and the most significant applications of 1D periodic, quasiperiodic, and disordered photonic structures. Concerning 1D disordered photonic structures, we particularly focus on 1D photonic structures in which the disorder is obtained in different ways: i) with a random sequence of high and low refractive index layers [45]; ii) with a random arrangement of a fixed number of high refractive layers between a fixed number of low refractive index layers [59,60]; iii) with a random variation of the thickness of the layers in a periodic structure [61–63].

## 2. Theoretical methods
*2.1. Brief summary of the theoretical methods to study photonic structures*
A wide variety of well-established theoretical methods to design and simulate the optical properties of photonic crystals is reported in literature. Most of the methods are exhaustively listed and discussed in the work of Prather et al. [64], as for example plane-wave expansion method (PWEM) [65–69]; finite-difference time-domain (FDTD) method [70,71] and finite element method [72,73].

In the particular case of one dimensional photonic structures, the transmission spectra can be simulated with different numerical tools, as finite element method [74], finite difference time domain methods [75]. The matrix methods are particularly interesting as they can be solved analytically, and we devote the following paragraph to introducing the transfer matrix method as a compelling tool to simulate the transmission spectra of 1D photonic crystals.

*2.2. Matrix methods for 1D photonic structures*

The matrix methods are very simple and versatile to simulate the optical properties of 1D photonic structures. Studies employing the scattering matrix method [76–78], impedance matrix method [79], and transfer matrix method [80–83] are reported.

For all the simulations presented in this manuscript we have employed the transfer matrix method. We consider a system (e.g. glass/multilayer/air) that is impinged by the light with normal incidence. The parameters related to air and substrate are just their refractive indexes, $n_0$ and $n_s$ respectively. With $E_m$ and $H_m$ the electric and magnetic fields in the glass, we can write the system that gives the electric and magnetic fields in air, $E_0$ and $H_0$:

$$\begin{bmatrix} E_0 \\ H_0 \end{bmatrix} = \prod_{j=1}^{t} M_j \begin{bmatrix} E_m \\ H_m \end{bmatrix} = \begin{bmatrix} m_{11} & m_{12} \\ m_{21} & m_{22} \end{bmatrix} \begin{bmatrix} E_m \\ H_m \end{bmatrix} \qquad (1)$$

with $t$ number of layers. $M_j$ is the characteristic matrix of each $j$ layer [$j$=(1,2,…,$t$)]

$$M_j = \begin{bmatrix} \cos(\phi_j) & -\frac{i}{p_j}\sin(\phi_j) \\ -ip_j\sin(\phi_j) & \cos(\phi_j) \end{bmatrix} \qquad (2)$$

$\phi_j$ is the phase variation of the light wave passing through the $j$th layer. For normal incidence $\phi_j = (2\pi/\lambda)n_j d_j$, where $n_j$ is the refractive index of the layer and $d_j$ its thickness. $p_j = \sqrt{\varepsilon_j/\mu_j}$ in transverse electric (TE) wave, while $q_j=1/p_j$ replace $p_j$ in transverse magnetic (TM) wave. At normal incidence the transmission spectra for TE and TM waves are the same.

The transmission coefficient can be written

$$t = \frac{2p_s}{(m_{11}+m_{12}p_0)p_s+(m_{21}+m_{22}p_0)} \qquad (3),$$

and the transmission

$$T = \frac{p_0}{p_s}|t|^2 \qquad (4);$$

the reflectivity coefficient is

$$r = \frac{(m_{11}+m_{12}p_0)p_s-(m_{21}+m_{22}p_0)}{(m_{11}+m_{12}p_0)p_s+(m_{21}+m_{22}p_0)} \qquad (5)$$

and the reflectivity

$$R = |r|^2 \qquad (6).$$

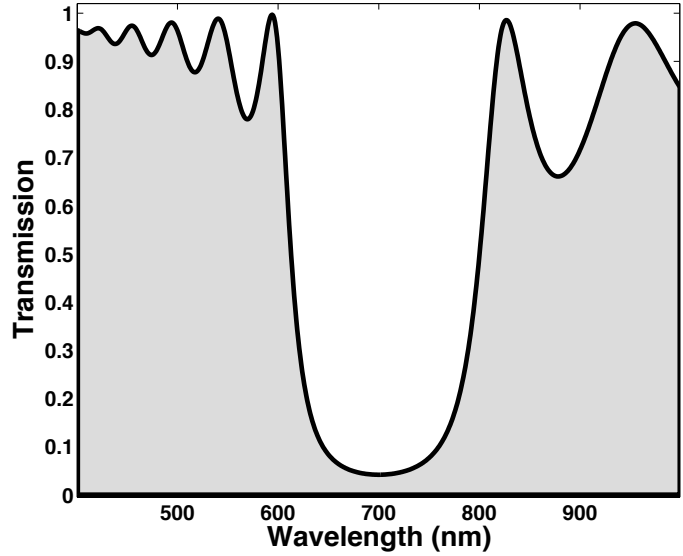

**Figure 4.** Example of the transmission spectrum of a periodic 1D photonic crystal (thickness of all the layers $d$ = 100 nm, high refractive index $n_h$ = 2, low refractive index $n_l$ = 1.46). In this work we call total transmission the grey area below the spectra (specifying the wavelength range).

In this article we refer to a quantity that we call normalized total transmission. The total transmission is the numerical integral of the transmission $[T(\lambda)]$ in the selected range of wavelengths (grey area in Figure 4, nm [400,1000]). The normalized total transmission is the total transmission normalized by a certain total transmission value in order to facilitate the comparison between the light transmission of different structures [e.g.: with $i$ different photonic structures, if we normalize all the total transmissions $T_t$ by the value of structure 3, the normalized total transmission $T_{nt}$ of the $i$th structure will be $T_{nt}^{(i)} = T_t^{(i)}/T_t^{(3)}$].

*2.2.1. Refractive indexes of the most important dielectric materials*
To study the transmission of a 1D photonic structure with two or more selected materials, it is proper to consider the dispersion of the refractive index of such materials. The dispersion of the refractive index can be described with the Sellmeier equation:

$$n^2 - 1 = \frac{A\lambda^2}{\lambda^2 - B^2} + \frac{C\lambda^2}{\lambda^2 - D^2} + \frac{E\lambda^2}{\lambda^2 - F^2} \qquad (7)$$

Many refractive index dispersions are reported in [84], with relative references. In Table 1 we report the coefficients of the Sellmeier equation of several materials that are usually employed for the fabrication of 1D photonic structures. We selected some inorganic materials and organic materials. The Sellmeier equations with these coefficients fit well the experimental data in specific wavelength ranges, that we reported for each material. Furthermore, we would like to stress that the reported Sellmeier equations refer to the real part of the refractive index. Thus, we do not consider the regions with absorption resonances. For example, we avoid to consider the UV region for PVK and for BDAVBi, where their absorption resonances occur since they are organic conjugated materials.

| Material | *A* | *B* | *C* | *D* | *E* | *F* | Ref. |
|---|---|---|---|---|---|---|---|
| Inorganic materials | | | | | | | |
| $SiO_2$ [a] | 0.6961663 | 0.0684043 | 0.4079426 | 0.1162414 | 0.8974794 | 9.896161 | [85,86] |
| $Al_2O_3$ [b] | 1.023798 | 0.0614482 | 1.058264 | 0.1106997 | 5.280792 | 17.92656 | [87] |
| $HfO_2$ [c] | 1.9558 | 0.15494 | 1.345 | 0.0634 | 10.41 | 27.12 | [88] |
| Y₂O₃ [d] | 2.578 | 0.1387 | 3.935 | 22.936 | - | - | [89] |
| ZrO₂ [e] | 1.347091 | 0.062543 | 2.117788 | 0.166739 | 9.452943 | 24.32057 | [90] |
| Si₃N₄ [f] | 3.0249 | 0.1353406 | 40314 | 1239.842 | - | - | [91] |
| Organic materials | | | | | | | |
| PVA [g] | 1.149 | 0.1234783 | - | - | - | - | [92] |
| PMMA [h] | 1.1819 | 0.011313 | - | - | - | - | [84,93] |
| PPO [i] | 0.0857 | 0.3183 | 0.5994 | 0.03294 | 0.7395 | 0.03288 | [94] |
| PVK [j] | 0.09788 | 0.3257 | 0.6901 | 0.1419 | 0.8513 | 0.1417 | [95,96] |
| CA [k] | 0.6481 | 0.0365 | 0.5224 | 0.1367 | 2.483 | 13.54 | [95,96] |
| HB-PVS [l] | 0.09938 | 0.3434 | 0.786 | 0.1742 | 0.8613 | 0.1758 | [97] |
| BDAVBi [m] | 0.3616 | 0.4273 | 1.542 | 0.1646 | 0.2339 | 0.1651 | [98] |

[a] range 210 – 6700 nm; [b] range 265.2 – 5577 nm; [c] range 365 – 5000 nm; [d] range 250 – 9600 nm; [e] range 361 – 5135 nm; [f] range 310 – 5504 nm; [g] polyvinyl alcohol (PVA), range 405 – 635 nm; [h] polymethyl methacrylate (PMMA), range 437 – 1052 nm; [i] poly(p-phenylene oxide) (PPO), fit of the experimental data in Ref. [94] in the range 400 – 1000 nm; [j] polyvinyl carbazole (PVK), fit of experimental data in Ref. [95] in the range 350 – 1000 nm; [k] cellulose acetate (CA), fit of the experimental data in Ref. [95] in the range 250 – 1000 nm; [l] hyperbranched polyvinylsulfide (HB-PVS), fit of the experimental data in Ref. [97] in the range 400 – 1500 nm; [m] 4,4'-Bis[4-(diphenylamino) styryl]biphenyl (BDAVBi) fit of the experimental data in Ref. [98] in the range 458 – 800 nm.

**Table 1.** Coefficients of the Sellmeier equation for different materials employed for the fabrication of 1D photonic structures. The references for each materials and the ranges of validity of the Sellmeier equations are reported.

The dispersion of the refractive index of $TiO_2$ (frequently used for the fabrication of 1D photonic crystals due to its high refractive index) needs to take into account several materials aspects, as for example crystalline phase. The refractive index of $TiO_2$ sputtered thin film in 1D photonic crystals can be expressed with the equation [99]:

$$n_{TiO_2}(\lambda) = \left(4.99 + \frac{1}{96.6\lambda^{1.1}} + \frac{1}{4.60\lambda^{1.95}}\right)^{1/2} \qquad (8)$$

Concerning plastic materials, an interesting work reports the Cauchy equations for several optical plastic materials [100] that can be employed to fabricate stratified 1D photonic structures. We would like to stress that, for polymeric materials, we should pay attention on the properties of the polymers investigated, as possible functional groups and average molecular weight. The Sellmeier equations reported here for polymers refer to the materials studied in the cited references. In particular, we have determined the coefficients for the Sellmeier equations of poly(p-phenylene oxide) (PPO), polyvinyl carbazole (PVK), cellulose acetate (CA), hyperbranched polyvinylsulfide (HB-PVS), and the molecule 4,4'-Bis[4-(diphenylamino) styryl]biphenyl (BDAVBi) by a fit of the experimental data (ellipsometric measurements) in Refs. [94–98].

Toccafondi et al. report the refractive index of the photochromic polymer p-DTE in the blue phase and in the transparent phase [101].

*Composite media*: In the case of layers made with composite materials, or with porous materials (thus, material and air), an effective dielectric constant (and, thus, the refractive index) can be described with several models, as the Lorentz-Lorenz formula [102]

$$\frac{\varepsilon_{eff}-1}{\varepsilon_{eff}+2} = f\frac{\varepsilon_1-1}{\varepsilon_1+2} + (1-f)\frac{\varepsilon_2-1}{\varepsilon_2+2} \qquad (9).$$

Otherwise, the dielectric constant can be described by the Maxwell-Garnett effective medium approximation [103–105]

$$\varepsilon_{eff} = \varepsilon_2 \frac{2(1-f)\varepsilon_2+(1+2f)\varepsilon_1}{2(2+f)\varepsilon_2+(1-f)\varepsilon_1} \qquad (10)$$

in both cases *f* is fraction of film volume filled by the material 1. An other type of effective medium approximation is the Bruggeman's model [106].

*Magneto-optical effects*: an external stimulus as the magnetic field can affect the refractive index. A phenomenon called Faraday rotation is a consequence of the refractive index difference for right and left circularly polarized light beams; such beams propagate through a material that is placed in a external quasi-static magnetic field ***B***, with light and magnetic field collinear. The wavelength dependent refractive index can be written [107]:

$$n_{R,L}(\lambda, B) = \sqrt{\varepsilon\mu} \pm \frac{\lambda V(\lambda) B}{2\pi} \qquad (11)$$

where *V* is the Verdet constant of the material and is also wavelength dependent. The material with one of the highest Verdet constant is TGG ($Tb_3Ga_5O_{12}$) [108–111].

### 3. 1D periodic photonic crystals
*3.1. Optical response of 1D periodic photonic crystals*

The optical properties of 1D photonic crystals are very well established and reported in many textbooks and articles, as for example [1–3,8,9,80,112]. Here, we just focus on the most important properties. The photonic band gap, the region in which light is not allowed to propagate through the medium, arises as a consequence of the 1D periodicity of the refractive index. In Figure 5 (top) we show the reflectivity, calculated with the transfer matrix method (the reflectivity is expressed in Equation 6), of a 1D photonic crystals made of 20 bilayers. In the bilayer, the higher refractive index is $n_h$=2, while the lower refractive index is $n_l$=1.55. The thickness of each layer is 650/4*n* nm.

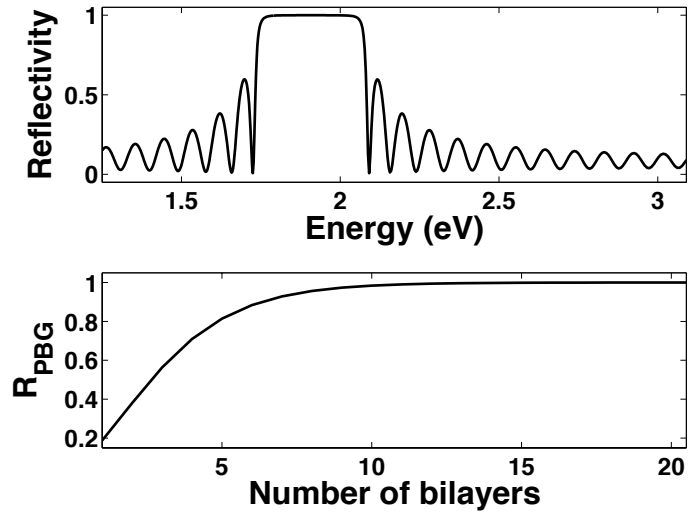

**Figure 5.** (top) Reflectivity of a 1D photonic crystal of 20 bilayers with $n_h$=2 ad $n_l$=1.55. The thickness of each layer is 650/4*n* nm. (bottom) Reflectivity at the center of the photonic band gap as a function of the number of bilayers.

In the case of Figure 5 the thickness *d* of each layer of the 1D photonic crystals is equal $\lambda_{PBG}/4n$, where $\lambda_{PBG}$ is the central wavelength of the first order of the photonic band gap and *n* is the refractive index of the layer. Thus we are in the lambda fourths condition. In this case, the intensity of the reflectivity is [6,80]

$$R = \left(\frac{1-\left(\frac{n_s}{n_0}\right)\left(\frac{n_h}{n_s}\right)^{2N}}{1+\left(\frac{n_s}{n_0}\right)\left(\frac{n_h}{n_s}\right)^{2N}}\right)^2 \tag{12}$$

where $n_l$ is the lower refractive index, $n_h$ is the higher refractive index, and $N$ is the number of bilayers. We show the trend of the reflectivity as a function of the number of bilayers in Figure 5 (bottom). Moreover, the band gap width (in energy) can be expressed in the following way [1,6]

$$\Delta E = \frac{4E_{PBG}}{\pi}\frac{n_h-n_l}{n_h+n_l} \tag{13}$$

In the case of the aforementioned 1D photonic crystal, ΔE is 309 meV.

*3.2. Fabrication techniques*
Various techniques have been employed to 1D photonic crystals where reproducible deposition of thin layers, is mandatory to achieve a high optical quality. Processes like ion implanting [113], sol–gel [114,115], electron-beam evaporation [116], Pulsed Laser Deposition [117,118], PECVD [119] and sputtering [120–123] can be successfully employed for the fabrication of microcavities based.
Anyway the deposition protocols have to careful be tailored on the particular materials employed and possible application of the fabricated system. More in detail semiconductor materials are successfully employed for the fabrication of 1D periodic photonic crystals but also dielectric materials could be used and recent works demonstrate as the addition of metal layers allow to combine 1D-PC with the surface plasmon resonance (SPR) to realize optical sensors [119].
Oxide-based dielectric materials are particularly suitable for fabricating photonic band gap (PBG) structures because they have wide transparency from the ultraviolet to the near infrared (NIR). Furthermore, oxide-based dielectric materials have good resistance to temperature, corrosion and radiation as well [120,124,125]. However, to reach high Q factor using dielectric material, where the refractive index difference between the different materials is not so high as for the semiconductor, the real time control of the deposition process is mandatory to allow a precise tailor of the deposition rate and obtain an enough good uniformity in thickness. Moreover, the increasing of the interest in shifting the rejected wavelengths in different zone ranging from visible to near infrared [125,126], requires an accurate design of the structures [127] and the definition of flexible experimental protocol capable to adapt itself to different materials and spectral range. A possible way to monitor the thicknesses of the processed film during the deposition procedure is represented by the quartz-crystal microbalance (QCM) which could be used for monitoring the growth-rate in physical vapour deposition and sputtering processes [128]. Sputtering methods are widely used in industrial process because high quality films can be obtained at low temperature substrates [129]. Moreover, it is also demonstrated as the rf sputtering is a suitable technique for fabrication of dielectric microcavities and it is a cheap and versatile technique to deposit alternating layers of different materials with controlled refractive index and thickness [122]. Finally, sputtering methods allow to process samples with a wide area, resulting in a relatively lower production cost per unit. With these advantages, as well as the possibility to incorporating QCM, rf-sputtering process is an extremely appropriate candidate to fabricate high quality and homogeneous 1-D photonic crystals.
For mesoporous materials, inorganic nanoparticles and polymers, the spin coating technique is very versatile and allows to obtain high optical quality 1D photonic crystals [6,94,130–132]. In the case of mesoporous materials and inorganic nanoparticles, a thermal annealing after each deposition (or each bilayer deposition, depending on the employed materials) is useful for the multilayer system fabrication [131,132]. In the case of polymers, a critical aspect is the solvent of the polymer solution. A judicious choice of the solvent is needed in order to avoid that the solvent of a polymer is going to dissolve the other, already deposited, polymer during the spin coating fabrication. Thus, a solvent is orthogonal when is a non-solvent for a specific polymer [133]. In literature couples of polymers, with corresponding orthogonal solvents are reported. A very interesting example is reported by Komikado et al. [134]: i) chlorobenzene has been selected because is a good solvent of PVK, and is an orthogonal solvent for cellulose acetate (CA); ii) diacetone alcohol (IUPAC name 4-Hydroxy-4-methylpentan-2-one) has been selected because is a good solvent for CA and is an orthogonal solvent for PVK. An other very interesting technique to obtain very high quality 1D multilayer photonic crystals with polymers is the co-extrusion process, as reported in Refs. [6,135–138]. Also organic semiconductors have been used to make a 1D photonic crystal with evaporation sources under vacuum [98].

*3.3. Applications*

Since 1D photonic crystals are versatile building blocks for myriads of applications, we can not give a complete overview in this review article. We present some examples of applications, as sensing, tunability upon an external stimulus, luminescence enhancement and lasing. Moreover, the employment of 1D photonic crystals in solar cells is briefly discussed.

*Sensing*: In Scheme 1 we depict the operation of a sensor based on a 1D photonic crystal. The presence of a certain analyte in the chamber is going to change either the effective refractive index or the thickness of the layers of the photonic structure, resulting in a shift of the photonic band gap.

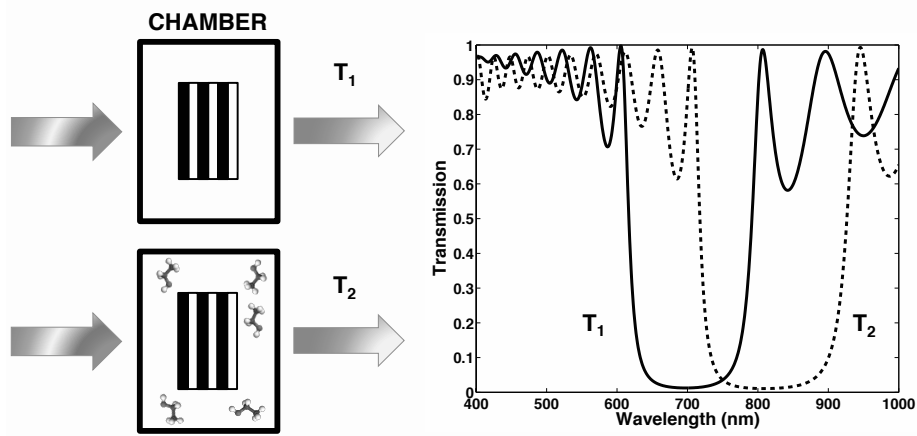

**Scheme 1.** Example of a sensor set-up with a 1D photonic crystal. The presence of a certain analyte in the chamber where the photonic crystal is placed is going to modify the photonic structure, in terms of layer refractive index or layer thickness. This results in a shift of the photonic band gap.

A photonic response to analyte requires, simply speaking, requires an interaction between the analyte and the materials employed for the fabrication of the photonic crystal. Porosity of the photonic crystal can increase the interaction allowing a high sensitivity of the analyte sensing. Moreover, the type of interaction is essentially important in terms of selectivity towards a particular analyte.

There are several reports on sensors based on 1D photonic crystals, fabricated with porous materials. Just to give few examples of these type of photonic crystals, Choi et al. report a one-dimensional photonic crystal composed by layers of mesoporous $SiO_2$ and mesoporous $TiO_2$ ,

demonstrating chemical sensing ability [139]. 1D photonic crystals using colloidal suspensions of functionalized mesoporous silica nanoparticles and titania sols have been studied for vapour sorption [140]. Clay Laponite based 1D photonic crystals show a structural colour that is sensitive to a wide range of analytes and can be employed as low cost sensing platforms [132,141]. 1D metal-organic framework photonic crystals have been suggested for detection of vapours [142]. Moreover, Bonifacio et al. proposed 1D porous photonic crystal for a "photonic nose" sensor platform for molecule and bacteria identification [143], and for water and food quality control [144].

Lova et al. [145] proposed a 1D photonic crystal made of cellulose acetate layers and ZnO nanoparticle loaded polystyrene layers. In this structure ZnO nanoparticles increase at the same time the effective refractive index of polystyrene layers and the free volume in the polystyrene matrix. The latter results in a high permeability to vapour, making the photonic crystal a good candidate for vapour sensing. In the work by Lova et al., a high sensitivity to toluene vapour.

We would like to mention that a general overview on the different ways to employ a 1D photonic crystal for chemical and biological sensing is given by Pavlichenko et al. [146] and by Lova et al. [94].

*Tunability*: The photonic band gap shift depicted in Scheme 1 can be seen in terms of tunability of the photonic band gap upon an external stimulus. The tunability can be exploited for the fabrication of displays. The stimulus can be either the presence of a chemical compound (as for a sensor) or an electromagnetic field.

A astonishing tunability of 575% (with a gap shifting from 350 nm to 1600 nm) has been demonstrated with a block-copolymer photonic gel, upon contact with a fluid reservoir [30]. Block-copolymer based 1D photonic crystals are also able to manifest an electrically-driven tunability, with the (PS-P2VP) copolymer [147] and the (PS-*b*-QP2VP) copolymer [148]. With a liquid crystal based 1D porous photonic crystal a shift of the photonic band gap of 8 nm with the application of a 8 V external voltage has been reported [149]. A similar performance has been shown with a Ag nanoparticle / $TiO_2$ nanoparticle 1D photonic crystal upon external voltage [150].

A very exhaustive review on electrically responsive photonic crystals, with many 1D examples, have been published by Nucara et al. [151].

*Luminescence enhancement*: One of the interesting features of the 1D microcavities is the possibility to enhance the luminescence, resonant with the cavity, when the defect layer is activated by a luminescent species. This is a general and significant property of photonic crystals and it has frequently been used to modulate the emission wavelength and enhance the radiative rate and intensity of luminescent objects [152,153]. When the cavity dimensions approach the wavelength of the emission, the density of electromagnetic states inside the cavity are strongly perturbed and can lead to significant enhancement of the luminescence quantum yield [154]. This enhancement is achieved by increasing the number of the localized modes coupled with the emitter [155,156].

Figure 6a shows the SEM image of a microcavity with an $Er^{3+}$-doped $SiO_2$ active layer inserted between two Bragg reflectors, each one constituted of ten pairs of $SiO_2/TiO_2$ layers. The dark regions correspond to the $SiO_2$ layer and the bright regions correspond to the $TiO_2$ layer. The substrate is located at the bottom of the images and the air on the top.

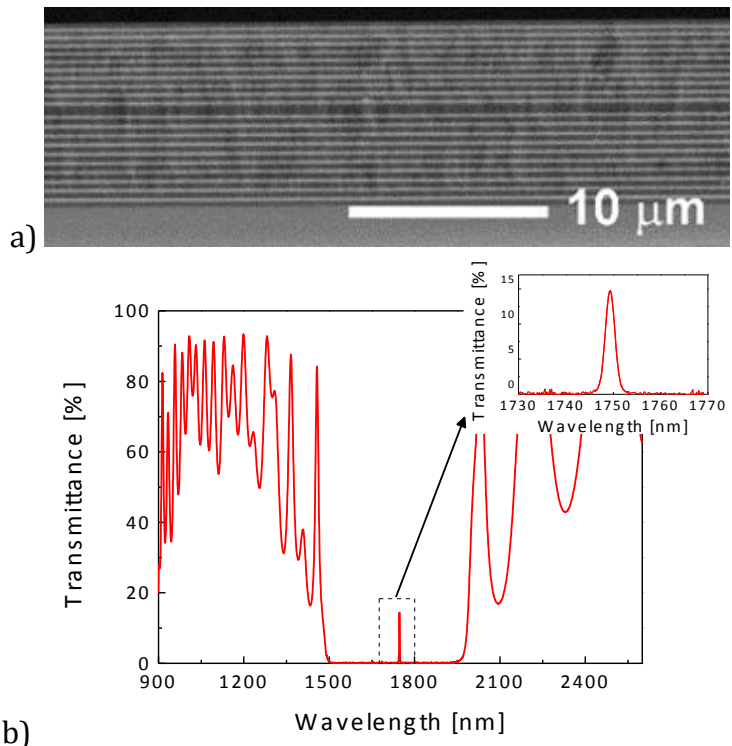

**Figure 6.** (a) SEM micrograph of a 1D microcavity fabricated by RF-sputtering. The $Er^{3+}$-doped $SiO_2$ active layer is inserted between two Bragg reflectors, each one constituted of ten pairs of $SiO_2/TiO_2$ layers. The bright and the dark areas correspond to $TiO_2$ and $SiO_2$ layers, respectively. The substrate is located on the bottom of the images and the air on the top. (b) Transmittance spectrum of the cavity with two Bragg reflectors, each one consisting of ten pairs of $SiO_2/TiO_2$ layers, in the region between 1000 and 2600 nm. The stop band range from 1490 to 1980 nm. The cavity resonance corresponds to the sharp maximum centred at 1.749 µm.

The sample was prepared by multi target rf sputtering technique [122] using silicon and silica substrates. The substrates were cleaned inside the rf sputtering deposition chamber by heating at 120 °C for 30' just before the deposition procedure. Sputtering deposition of the films was performed by sputtering alternatingly a 15×5 cm² titania target and a 15×5 cm² silica target. For the defect layer a 15×5 cm² silica target, on which metallic erbium pieces were placed, was employed. The deposition time, necessary to reach the appropriate thickness of the Bragg mirror layers, was 2 h 30 min for titania target and 1 h 20 min for silica target. The deposition time necessary to reach the appropriate thickness of the silica defect layer, to obtain cavity resonance centred at 1.749 µm, was 2 h 40 min. The residual pressure, before the deposition, was about $1.1\times10^{-6}$ mbar. During the deposition process, the substrates were not heated and the temperature of the sample holder during the deposition was 30 °C. The sputtering occurred with an Ar gas pressure of $5.4\times10^{-3}$ mbar; the applied rf power was 150 W and 130 W and the reflected powers 0 W for silica and titania targets, respectively. Particular attention was paid to the reproducibility of the single layers. To monitor the thickness of the layers during the deposition, two quartz microbalances Veeco instruments thickness monitor model QM 311, faced on the two targets were employed. Thickness monitor was calibrated for the two kinds of materials by a long deposition process (24h of deposition) and by directly measuring the thickness of the deposited layer by an m-line apparatus [122]. The final resolution on the effective thickness obtained by this quartz microbalance is about 4 Å. The $Er^{3+}$ content in the active layer is about 0.6 ± 0.1 mol%. The NIR transmittance spectrum, measured at zero degree of incident angle, is reported in figure 6b and shows the stop band from 1490 to 1980 nm. A sharp peak in the transmittance spectrum appears at 1749 nm, as presented in the inset of figure 6b. It corresponds to the cavity resonance wavelength related to the half wave layer inserted between the two Bragg mirrors. The full width at half maximum of the resonance is 1.97 nm, corresponding to a quality factor of the cavity, Q, of about 890.

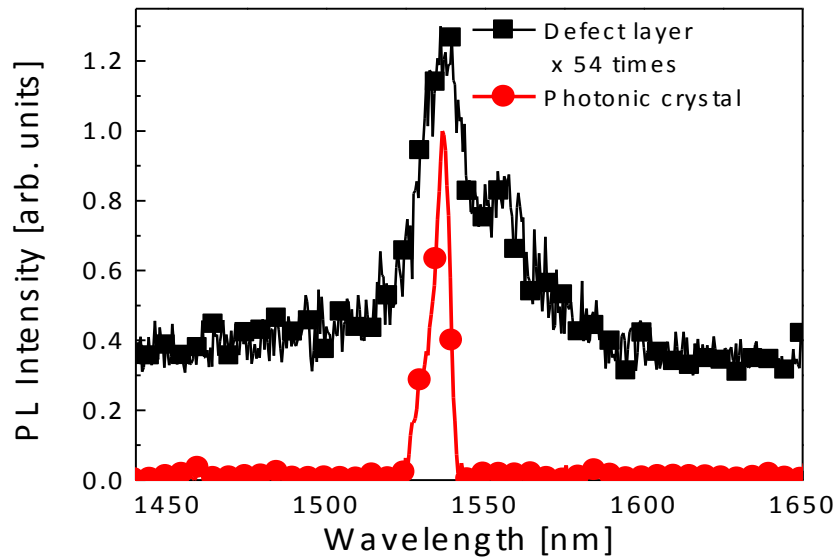

**Figure 7.** $^4I_{13/2} \rightarrow {}^4I_{15/2}$ photoluminescence spectra of the cavity activated by $Er^{3+}$ ion in 1D photonic crystal (●) and of the single $Er^{3+}$-doped $SiO_2$ active layer with first Bragg mirror (■). The light is recorded at 50° from the normal on the samples upon excitation at 514.5 nm.

Figure 7 shows the luminescence from the cavity and from the $Er^{3+}$-doped single $SiO_2$ layer with one Bragg reflector. In order to get a correct comparison, a specific procedure, assuring that the only variation is constituted by the cavity effect, was employed as reported in detail in [122]. Both the cavity and the $Er^{3+}$-doped single $SiO_2$ layer with first Bragg reflector were excited with the 514.5 nm line of an $Ar^+$ ion laser with an excitation power of 180 mW. The erbium emission from the reference sample is centered at 1538 nm with a FWHM of 29 nm and exhibits the characteristic shape of $Er^{3+}$ ion in silica glass [157]. The peak luminescence intensity of $Er^{3+}$ ions is enhanced by a factor 54, in respect to that detected for the reference at the corresponding wavelength. The $Er^{3+}$ $^4I_{13/2} \rightarrow {}^4I_{15/2}$ emission line shape is strongly narrowed by the cavity and exhibits a full width at half maximum of 5 ± 0.5 nm. The $Er^{3+}$ emission is enhanced when the wavelength corresponds to the cavity resonant mode and weakened for the others emission wavelengths depending on the number of the localized modes coupled with the erbium ion in the defect layer.

When the spontaneous emission of the emitter, embedded in the defect layer of a 1D photonic crystal, is strongly enhanced, the possibility of low threshold lasing could take place. In this regard, there have been considerable efforts to fabricate photonic band gap laser devices either as a distributed feedback lasing at band edge frequencies [158] or as a defect-mode lasing at localized defect mode frequencies [159]. In particular, due to the relative simplicity of fabrication, one dimensional 1D photonic crystal laser devices have been extensively studied. The more effective devices are fabricated by organic/ inorganic hybrid 1D photonic crystal and some years ago organic laser dyes as a gain medium was used to demonstrate a low threshold defect-mode lasing action [160]. It is demonstrated low threshold defect-mode lasing action in a one dimensional microcavity constituted by two Bragg reflectors, each one constituted of ten pairs of SiO$_2$/TiO$_2$ layers, with a defect layer based on poly-laurylmethacrylate matrix containing CdSe@Cd$_{0.5}$Zn$_{0.5}$S quantum dots [161]. The defect-mode laser structure is based on photophysical properties of the gain medium, and in particular in order to reflect the light at around 650 nm wavelength. The defect mode laser structure was optically pumped at 514.5 nm and the luminescence spectrum in the case of 2 mW of excitation power is shown in Figure 8. Experimental details are given in Ref. [161]. The spectrum shows some narrow peaks in the low energy region followed in the high frequency region by a typical comb structure superimposed to the broad band assigned to the spontaneous emission. We can conclude that the spontaneous emission in the low energy region, around the localized defect modes, is enhanced by a factor proportional to the density of states at those frequencies leading to low threshold lasing.

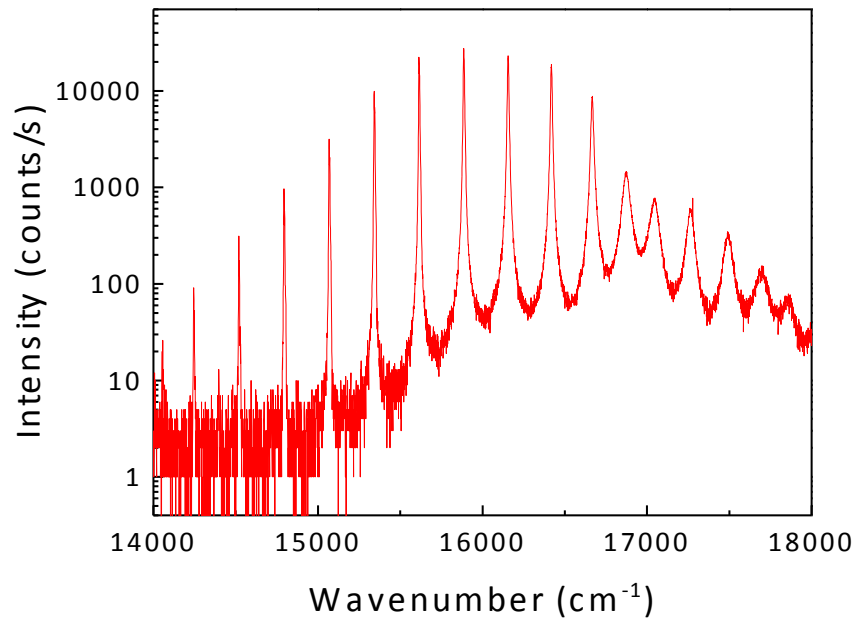

**Figure 8.** Luminescence spectrum obtained exciting at 514.5 nm with 2 mW, the SiO$_2$ /TiO$_2$ 1D microcavity, fabricated by RF sputtering, with a defect layer constituted by a poly-laurylmethacrylate matrix containing CdSe@CdZn$_{0.5}$S quantum dots.

The measurements were taken with a 2 cm$^{-1}$ step and different laser power, and the intensity of the 15803 cm$^{-1}$ peak was plotted as a function of the pump power; these data were then fitted via a linear function considering the points on the graph obtained at pump power above 0.5 mW. The fit line reaches the X-axis at a value of about 0.5 mW. This kind of not completely linear dependence could be due to presence of coherent emission from the sample. However, to find the proof of laser emission some more evidence had to be gathered.
As one can see in figure 9 the peak intensity can be approximately described via a linear dependence on the pump power up to 4 mW. It is important to note that in this range of pump power values, even at low power, the shape of the spectra does not change.

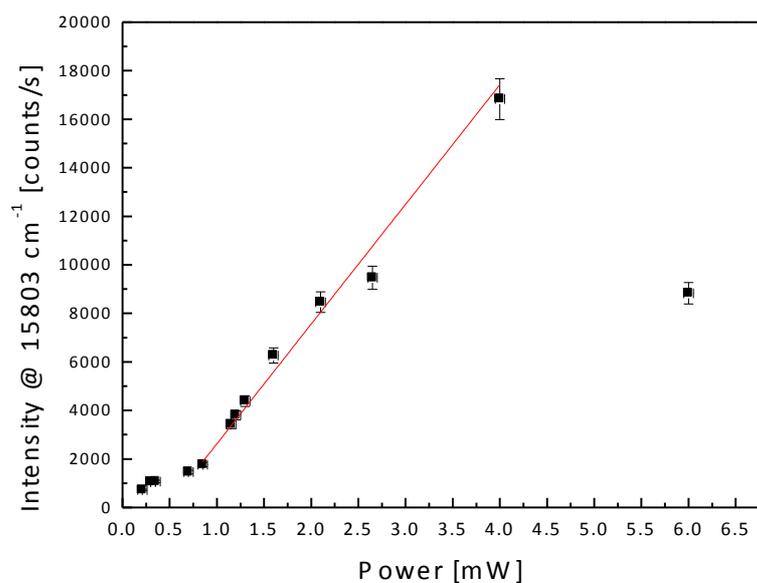

**Figure 9.** Intensity of the luminescence measured at 15803 cm$^{-1}$ at different pump power up to 6 mW focusing the excitation laser beam on the sample and detecting the luminescence at 2° with a solid angle of 7·10$^{-4}$. The line is the result of the linear fit on all the points.

A different type of luminescence enhancement is the enhancement of the emission of a organic light emitting transistor. With a proper design of the transistor, in which the 1D photonic crystal acts also as gate dielectric between the gate contact and the active material of the transistor, the emission can be modulated and enhanced [162,163].

*Lasers*: Lasing has been demonstrated with dye-doped polymeric microcavities [164] and all-polymeric microcavities [165]. The inclusion of an optically active material as a dye or a polymer in a photonic crystal leads to a laser called distributed feedback (DFB) laser that oscillate on a pair of wavelengths, corresponding to the edges of the photonic band gap. Such behaviour is deeply described by Kogelnik and Shank [166] and Dowling et al. [167].
Many examples of organic semiconductor DBF lasers are reported in Samuel and Turnbull [168], With polymeric materials as polyvinyl carbazole and cellulose acetate it has been demonstrated the realization of a DFB laser [134,169], also on a flexible substrate [170]. Furthermore, DFB lasers based on metal oxide nanoparticle photonic crystals have been realized, showing laser emission upon one-photon [171] and two-photon pumping [172].

*Solar cells*: The role of a photonic crystal in a photovoltaic device is either the localization or the re-injection of the photons, that are not absorbed by the active photovoltaic material, back to the device where charge separation takes place [173,174]. There are several seminal works envisaging the employment of a 1D photonic crystal to enhance the performance of a solar cell [175–178].
Among the most interesting achievements in the field, we would like to mention one of the first demonstrations of a power-conversion efficiency improvement in a dye sensitized solar cell (DSSC) with a 1D porous photonic crystal [179], the fabrication of a transparent 1D photonic crystal based DSSC [180], the integration of a porous photonic crystal in a organometal halide perovskite solar cell [181], the integration of 1D photonic crystal in a semitransparent polymer solar cell [182,183].

## 4. 1D photonic quasicrystals

*4.1. Optical response of 1D photonic quasicrystals*

Quasicrystals are deterministic aperiodic structures that do not have translational symmetry, but they show long range order and Bragg diffraction [17,184]. 1D quasicrystals are made by arranging a stack of two different materials A and B according to a deterministic generation scheme [11,13].

A very exhaustive review by Steurer and Sutter-Widmer [19] reports several types of 1D photonic quasicrystals and some significant studies on such structures: Fibonacci sequence [12,185,186]; Thue-Morse sequence [187,188]; Period-doubling sequence [189,190]; Rudin-Shapiro sequence [191]; Cantor sequence [192,193]; modulated structure [194,195].

A way to describe the Fibonacci sequence, based on the two alphabet letter A and B (corresponding to the two materials), and substitution rule σ(A)=AB and σ(B)=A, can be written as

$$\sigma: \begin{pmatrix} A \\ B \end{pmatrix} \rightarrow \begin{pmatrix} 1 & 1 \\ 1 & 0 \end{pmatrix} \begin{pmatrix} A \\ B \end{pmatrix} = \begin{pmatrix} AB \\ A \end{pmatrix} \tag{14}$$

where $S = \begin{pmatrix} 1 & 1 \\ 1 & 0 \end{pmatrix}$ is the substitution matrix [19]. An other way to describe the Fibonacci sequence follows the generation scheme [11]:

$$S_{j+1} = \{S_{j-1}S_j\} \text{ for } j \geq 1$$
$$\text{with } S_0 = \{B\} \text{ and } S_0 = \{A\} \tag{15}$$

In Figure 10 we show the transmission spectra of three different types of quasicrystals: a Fibonacci sequence of 21 layers (top); a Thue-Morse sequence of 32 layers (center); a Rudin-Shapiro sequence of 32 layers (bottom). The refractive indexes of the two materials are $n_A$ = 2.3 and $n_B$ = 1.8, while the thickness of each layer is $125/n_{A,B}$ nm (for the A and B materials, respectively).

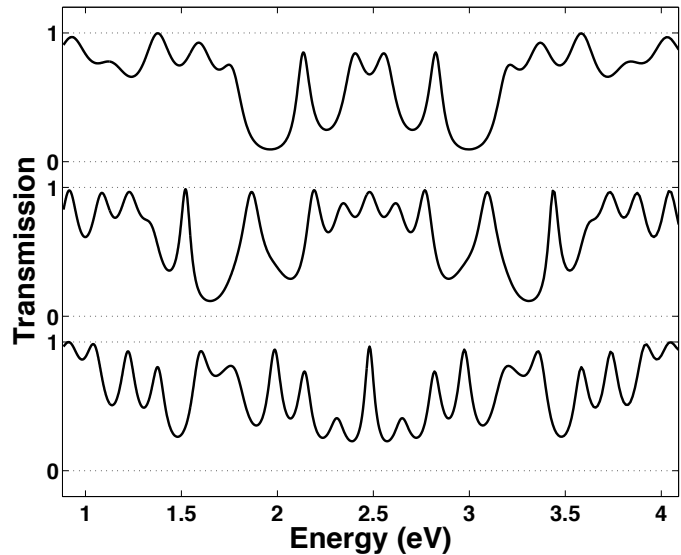

**Figure 10.** (top) Transmission spectrum of a 1D photonic Fibonacci quasicrystal made with 21 layers, following the sequence ABAABABAABAABABAABABA [19]. (center) Transmission spectrum of a 1D photonic Thue-Morse quasicrystal made with 32 layers (ABBABAABBAABABBABAABABBAABBABAAB) [19]. (bottom) Transmission spectrum of a 1D photonic Rudin-Shapiro quasicrystal made with 32 layers (AAABAABAAAABBBABAAABAABABBBAAABA) [19]. The refractive indexes

of the two materials A and B constituting the photonic quasicrystal are $n_A$ = 2.3 and $n_B$ = 1.8.

A new type of quasicrystal ha been designed and has been called Octonacci crystal [196–198]. Its generation scheme can be written as:

$$S_n = S_{n-1}S_{n-2}S_{n-1} \qquad (16)$$

for $n \geq 3$, with $S_1$ = A and $S_2$ = B.

*4.2. Fabrication techniques*
Gellermann et al. fabricated a $SiO_2/TiO_2$ 1D photonic Fibonacci quasicrystal with electron-gun evaporation at a pressure of $10^{-8}$ bar [186]. Dal Negro et al. [13] have realized Fibonacci multilayers made of porous silicon by electrochemical dissolution procedure [199,200]. Sahel et al. have employed radiofrequency magnetron sputtering to fabricate a Cantor sequence made with silicon and silicon dioxide [201], while Lusk and Placido have used microwave assisted direct current magnetron sputtering to fabricate a Fibonacci sequence [202]. Howkeye and Brett have fabricated with glancing angle deposition a Thue-Morse multilayer made of titania [203].

*4.3. Applications*
The possibility to achieve light pulse compression with a Cantor filter has been proposed by Garzia et al. [204] and Cojocaru [205]. Rea et al. have experimentally demonstrated photoluminescence enhancement of graphene oxide nanosheets when infiltrated in a porous silicon based Thue-Morse sequence [206].

**5. Disordered 1D photonic structures**
*5.1. Optical response of disordered 1D photonic structures*
The disorder in a 1D photonic structure can be induced in many ways. We can play with: i) the thickness of each layer, that can either be a random number or follows a specific distribution of layers; ii) the number of materials, to have materials C, D, E ... substituting the materials A and B in the photonic crystal; iii) the permutation in the arrangement of the materials in the photonic structures. In the latter (and in the case of two materials), we can decide the proportion of the two materials in the sequence (e.g. 20% of material A and 80% of material B) and arrange the layers by permutating them. Otherwise, the sequence can be arranged by a 50% probability to have either material A or material B (of course, in the thermodynamic limit of this system, we will reach a 50/50 proportion of the two materials).

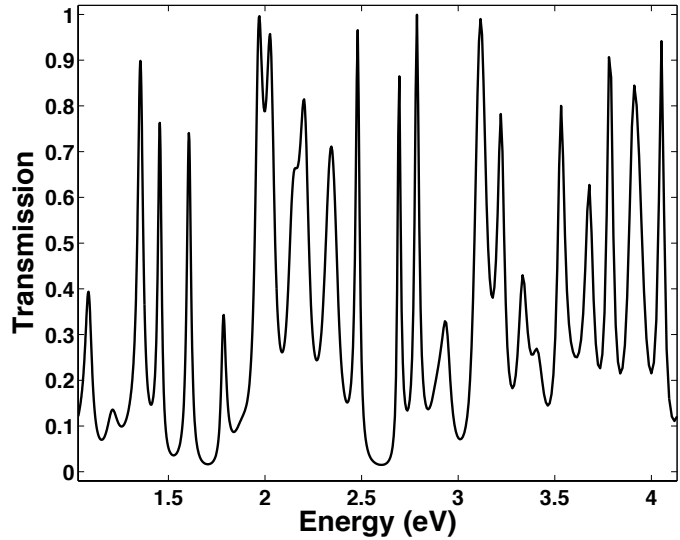

**Figure 11.** Transmission spectrum of a 1D disordered photonic structure with 28 layers. The thickness of each *i*th layer is a random number between 80 nm and 120 nm. The refractive index of each *i*th layer is a random number between 1.3 and 3.5.

In Figure 11 we show the transmission spectrum of 1D random structure in which all the layers have different thickness (in the range 80 – 120 nm) and a different refractive index (in the range 1.3 – 3.5). This thickness and "compositional" disorder leads to a spectrum with many transmission peaks and valleys.

In the following sub-paragraphs we will discuss the possibility to correlate the arrangement of 1D disordered photonic structures with their optical properties in terms of total transmission (Figure 1). First, we describe structures in which the high refractive index layers are aggregated in clusters and the homogeneity of the structure is quantified by the Shannon index. Second, we describe structures in which we control the distribution of high refractive index layer clusters by following specific behaviours, i.e. a power-law distribution and a uniform distribution.

*5.1.1. The Shannon index to quantify the homogeneity of 1D photonic structures*
To quantify the homogeneity of a studied structure we can employ the Shannon index (or Shannon-Wiener), also called Shannon entropy, a diversity index measure that reflects how many different types (such as species) there are in a dataset, and simultaneously takes into account how evenly the basic entities (such as individuals) are distributed among those types. Claude Shannon elaborated the formalism in 1948 to assess the entropy in strings of text [207]. The Shannon index is widely used in different fields of science, such as information theory [208,209], ecology [210–212], microbiology [213], statistics [214], statistical mechanics and physics [215].

In optical and condensed matter physics, researchers have used the Shannon entropy variation to assess the stability of a single photon based quantum cryptography protocol [216], to determine the probability of electronic charge distribution between the atoms of a benzene ring (also named aromaticity measure) [217], and to evaluate the electron localization in a molecular system [218]. Moreover, a parallelism in condensed matter physics among information lattice and subgroup lattice has been reported [219].
In two dimensional photonic structures, it has been demonstrated the normalized total transmission $T_{nt}$ can be related to the Shannon index $H'$ approximately with this simple linear behaviour

$$T_{nt} \cong \frac{H'}{2} + \frac{1}{2} \qquad (17)$$

that shows an increase of the normalized total transmission by increasing the Shannon index [220,221].

We consider here a 1D photonic crystal made by alternating layer of titanium dioxide and silicon dioxide. For titanium dioxide we consider a constant refractive index $n_T$ = 2.45, while for silicon dioxide we consider a constant refractive index $n_S$ = 1.46. The thickness of the titanium dioxide layer is 75 nm, while the thickness of the silicon dioxide one is 225 nm, i.e. three layers of 75 nm. In this way we can also say that, in the lattice unit of 300 nm, we have one 75 nm layer of titanium dioxide and three 75 nm layers of silicon dioxide. Such photonic crystal is depicted in Figure 12 (top).

We show in this work the possibility to correlate the distribution of the layers in the 1D photonic structure to the Shannon index value [207].

The Shannon index is defined as

$$H' = \frac{-\sum_{j=1}^{S} p_j \log p_j}{\log(s)} \qquad (18).$$

In the equation $p_j$ is proportion of the *j*th species in the structure and *s* is the number of the different species. The division by *log(s)* normalizes the index in the range (0,1). We can compute *H'* in the 1D photonic structure by dividing the structure length in a certain number of *s* linear sub-units. We consider the titanium dioxide as the *j*th species under study. The fraction of the titanium dioxide layers belonging to each sub-unit denotes the proportion $p_j$ in the Equation (6). *H'* is maximum when all sub-units contain the same number of titanium dioxide layers (maximum homogeneity of the medium). Instead, when all the layers are in only one sub-unit, *H'* becomes the minimum and such structure is the most aggregated (minimum homogeneity of the medium).

As mentioned above, in this specific study we analyze 1D photonic structures with 64 layers, 16 are made of titanium dioxide and 48 are made of silicon dioxide. By splitting the photonic structure in 16 sub-units, the most uniform (the most homogeneous, i.e. with *H'* = 1) structure is the one having a $TiO_2$ layer in each sub-unit. Contrariwise, since each sub-unit can contains up to four layers, the most aggregated linear configuration possible is a structure where four sub-units enclose four $TiO_2$ layers each. For the topology of studied structure, the aforementioned one is the most non-homogeneous structure, corresponding to *H'* = 0.5 (Figure 12, bottom). In Reference [59] it is reported a study on 1D photonic structures with 11 different grades of homogeneity, corresponding to Shannon indexes *H'* ranging in the interval (0.5, 1).

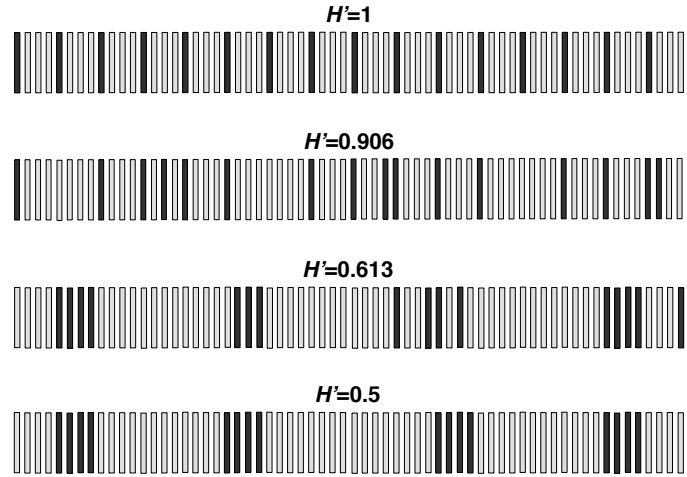

**Figure 12.** Schematic representation of 1D photonic structures with four different homogeneities. The higher structure with $H'=1$ is the most homogeneus; the lower structure with $H'=0.5$ is the most aggregated and less homogeneus.

Here we study 5000 different permutations of the 1D photonic structure with the Shannon index $H'$ ranging in the same interval (0.5, 1). In Figure 13 we report the normalized total transmission (in the range 650 – 1350 nm) of these structures as a function of their Shannon index. The total transmission has been normalized by the one of the periodic photonic crystal, having $H' = 1$ (star in Figure 13).

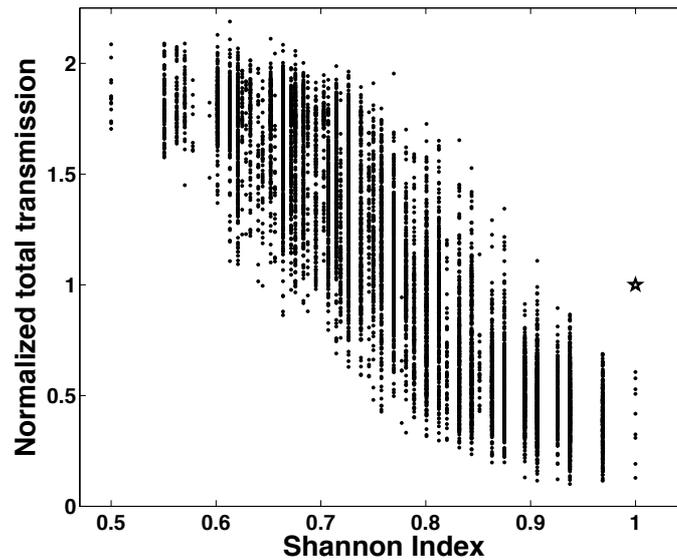

**Figure 13.** Normalized total transmission (normalized by the total transmission of the ordered photonic crystal) of $SiO_2/TiO_2$ 1D photonic structures with different homogeneity [59].

With many permutations of the 1D photonic structure, with respect to study in Reference [59], we have a more clear picture of the correlation between the normalized total transmission and the Shannon index. In particular, we observe a monotonic decrease of the total transmission as a function of the Shannon index, up to $H' = 1$. The periodic photonic crystal, with $H' = 1$, has a higher total transmission with respect to the other structures with $H' = 1$. The latter can be simply explained: while the periodic crystal has 16 unit cells (corresponding to the sub-units of Equation 1) that have the configuration of the layers $TiO_2$-$SiO_2$-$SiO_2$-$SiO_2$, the other structures may have some unit cells with a configuration $SiO_2$-$TiO_2$-

SiO$_2$-SiO$_2$, or SiO$_2$-SiO$_2$-TiO$_2$-SiO$_2$, or SiO$_2$-TiO$_2$-SiO$_2$-TiO$_2$. The Shannon index is always 1, but the periodicity of the structure is affected, consequently affecting the total transmission of the structure. It is interesting that the periodic photonic crystal, among the structures having $H' = 1$, is the one with the highest total transmission. This is because the periodic photonic crystal show a low transmission in the band gap region and high transmission in the other regions, while the other structures show some additional transmission valleys in the whole studied spectral interval (650 – 1350 nm), resulting in a lower total transmission (see Figure A1 in the Appendix).

*5.1.2. Power law distribution of high refractive index clusters in 1D photonic structures*
In some recent works [222,223] on the modelling of light transport in Levy glasses, the light absorption has been correlated to the power law of the step-length distribution of the light mean free path in the medium. The influence of the power law distribution, of the light free path in the medium, on the degree of superdiffusivity of these materials is reported in these studies. Generally, a power law distribution of the refractive elements in the sample is made to generate a heavily-tailed step length distribution for the random walk of light, expected to result in superdiffusion [222,224].

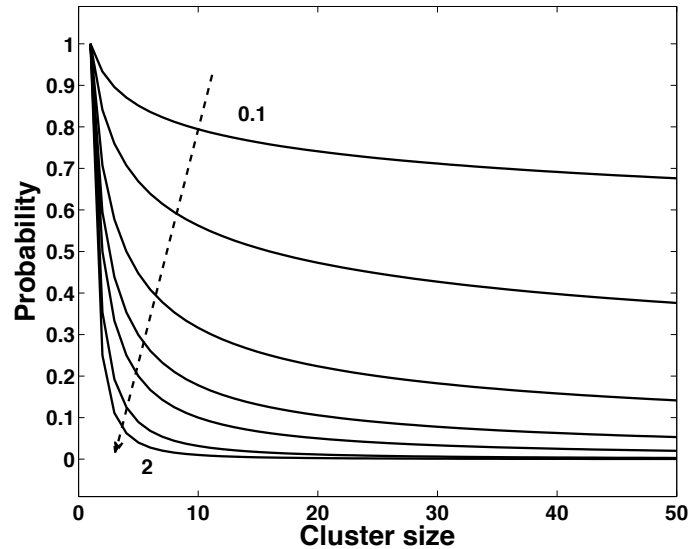

**Figure 14.** Power law distribution with different *a* value: 0.1, 0.25, 0.5, 0.75, 1, 1.5, 2. The dashed line highlights the increase of *a* along the different function.

In this work, we studied 1D photonic structures made of 360 high refractive index layers ($n_1 = 1.6$) with thickness $d_1 = 70$ nm and 2520 low refractive index layers ($n_2 = 1.4$) with thickness $d_2 = 80$ nm. The optical length $nd$ is 112 nm for all the layers. Moreover, the ratio between the high and low refractive index layers is 1/8. The unit cell of the engineered structure is thus composed by 1 high refractive index layer (we call it H) and 7 low refractive index layers (we call them L). The photonic structure has in this way 360 unit cells and can be simply schematized as $(H_1L_7)_{360}$.

The disordered 1D photonic structures studied here are characterized by the aggregation of the high refractive index ($n_1$) layers. In particular the cluster size distribution follows the power law:

$$f(x) = x^{-a} \quad , \tag{19}$$

in which *x* indicates the size of the cluster and *a* the exponent of the power law. For sake of clarity, we truncated the distribution in order to have clusters with a size that spans from 1 to 50. Examples of truncated distributions, for different exponent *a*, are reported in Figure 14. With *a* = 0, the distribution becomes a uniform distribution where all cluster sizes are equally probable. By increasing the value of *a* the probability for a large cluster to occur in the photonic structure is very small. In the assembly of the structure, the high refractive index layer clusters were randomly distributed within the 2520 low refractive index layers. For each exponent *a*, we have simulated the transmission spectra of 1000 different permutations of the high refractive index layer cluster within the low refractive index layers, in order to avoid any correlation between the light transmitted by the photonic structure and a particular cluster arrangement. We have simulated the transmission spectra as a function of the wavelength with a step of 0.1 nm.

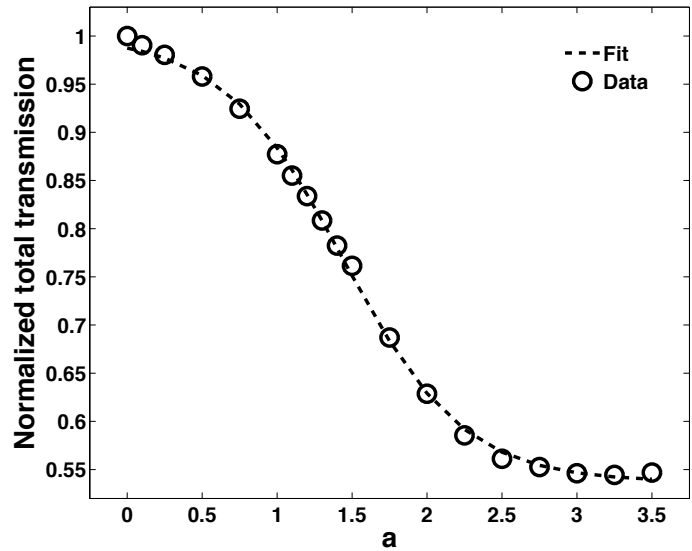

**Figure 15.** Normalized total transmission as a function of *a* for refractive index contrast Δn = 0.2 (trends for different Δn are reported in Ref. [225]). Circles are the experimental data while dashed line is the sigmoid function as reported in Equation 8.

We found that the total transmission, in the range of the photonic band gap of the structure $(H_1L_7)_{360}$, follows a characteristic trend as a function of the exponent *a* of the power law distribution of clusters. In Figure 15 such trend is reported, normalizing the total transmission by the value related to *a* = 0. In particular, the trend follows a sigmoidal function. By studying the trend for different refractive index contrast, we found a sigmoid equation:

$$T_{nt}(a) = 1 - \frac{1.42 - 0.3455\Delta n^{-0.6365}}{1 + e^{-(0.70\Delta n + 2.33)(a + \Delta n - 1.63)}} \qquad (20)$$

The fit is very interesting because has only one parameter, i.e. the refractive index contrast.

*5.1.3. Uniform distribution of high refractive index clusters in 1D photonic structures*
As in the experiment in Paragraph 5.1.2, in this study we have engineered 1D photonic structures in which clusters of high refractive index layers are arranged between low refractive index layers. Here, we chose different numbers of low refractive layers. As reported in Figure 16 (where only the first 24 layers of the structure are depicted), a unit cell contains *m* layers, one high refractive index layer and *m* – 1 low refractive index layers. 1/*m* thus indicates the ratio between high and low refractive index layers in the unit cell (and, consequently, in the photonic structure). We used 100 high refractive index layers, and 100 low refractive index layers for $m = 2$, 300 low refractive index layers for $m = 4$, 500 low refractive index layers for $m = 6$, and so on.

In order to realize the disordered structures, we have aggregated the high refractive index in clusters, with the size of such clusters that spans from 1 to a value, $k_{max}$, indicating the maximum cluster size (Figure 16 reports three examples). In this way, by increasing $k_{max}$ we increased the possibility to have large high refractive index layer clusters in the photonic structure. The clusters are then randomly arranged between the low refractive index layers.

We underline that all the studied photonic structures have the same number of high and low refractive index elements, but differ in the cluster arrangement (that is related to the homogeneity of the structure). More details on the studied photonic structures are reported in Ref. [82].

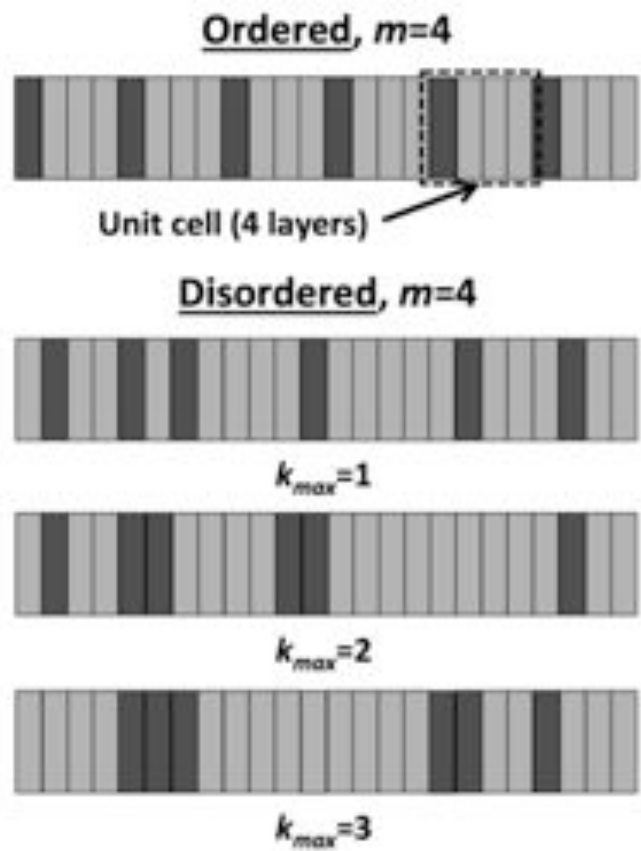

**Figure 16.** 1D photonic structures with different types of disorder.

We study for all the structures the total transmission, with a step of 1 nm, in the spectral interval corresponding to the photonic band gap of the periodic photonic crystal (e.g. in the case of $m = 4$, we calculate the total transmission in the spectral interval corresponding to the band of a periodic crystal with 100 unit cells, in which the unit cell contains 1 high refractive index layer followed by 3 low refractive index layers).

In Figure 17 we show the total transmission of the disordered photonic structures as a function of the maximum cluster size $k_{max}$. The total transmission trends are normalized by their maximum and each point depicted corresponds to the mean value of the total transmission of 1000 different permutations of the high refractive index layer clusters among the low refractive index layer clusters.

The trends have a characteristic behaviour: a valley in normalized total transmission followed by an increase with remarkable oscillations.

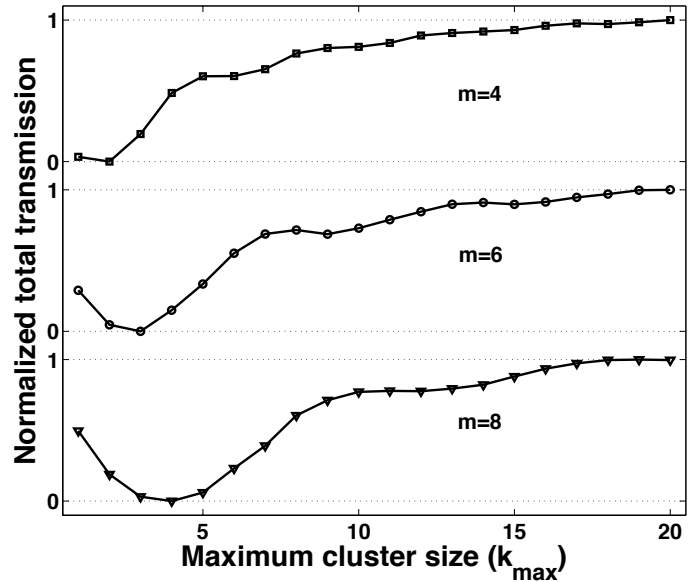

**Figure 17.** Normalized total transmission as a function of the maximum cluster size (trends for more values of *m* are reported in Ref. [82]).

Focusing the attention on the photonic structure with *m* = 8 (where valley and oscillations are very clearly observable), in Figure 18 we report the trend for different refractive index contrast Δn. The pattern holds for three different Δn, but the intensity of the valley increase with Δn.

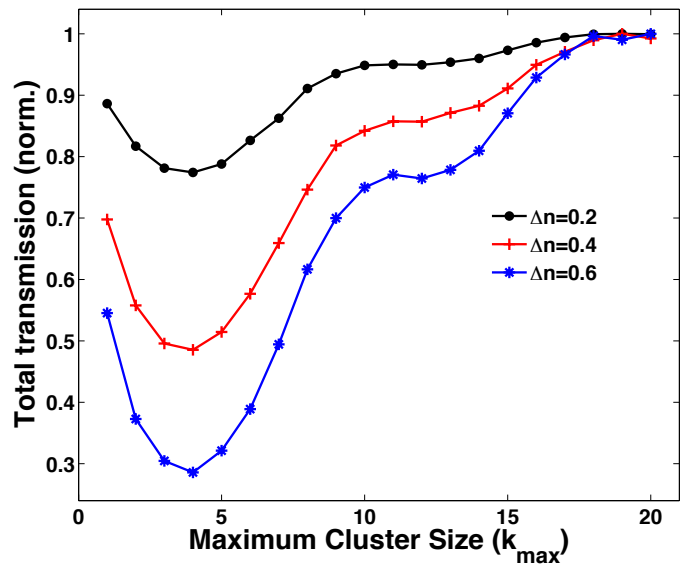

**Figure 18.** Total transmission (normalized to the maximum of the curve) as a function of the cluster size $k_{max}$ for three different refractive index contrast (with *m* = 8). The black curve corresponds to $n_1$ = 1.4 and $n_2$ = 1.6; the red curve corresponds to $n_1$ = 1.4 and $n_2$ = 1.8; the blue curve corresponds to $n_1$ = 1.4 and $n_2$ = 2.0.

An additional evidence of the robustness of the trend is shown in Figure 19 where the trend of the normalized total transmission is studied for the same difference $n_2 - n_1$, but with different refractive contrast. We do not observed variations in the trend (apart from the magnitude of the transmission valley and the oscillations as in Figure 18).

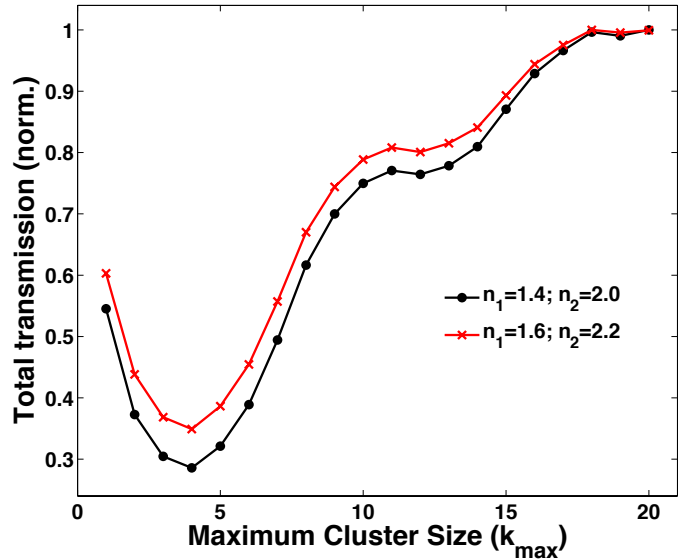

**Figure 19.** Total transmission (normalized to the maximum of the curve) as a function of the cluster size $k_{max}$ for the same value of $n_2 - n_1$, but for different refractive indexes (indicated in the legend).

The trend can be affected by very invasive changes in the disordered photonic structures, as the introduction of defects, e.g. layers of a third material with a different refractive index $n_3$. The third material layers (with $n_1 = 2.2$) exchange the low refractive index layers (with $n_1 = 1.4$).

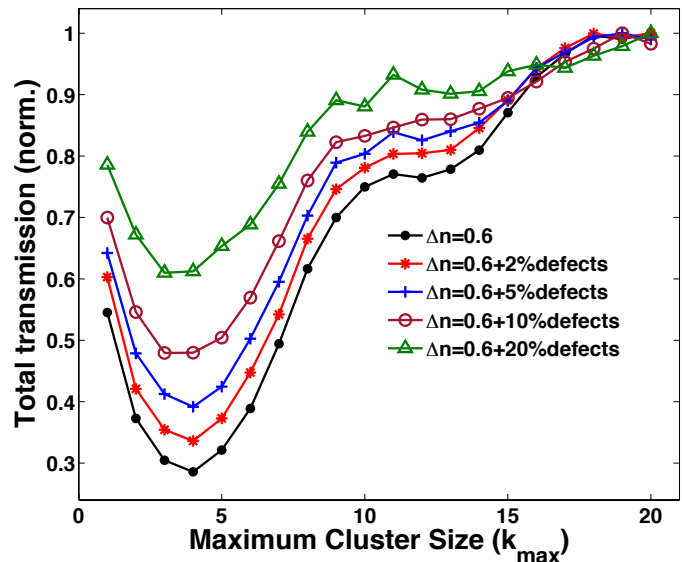

**Figure 20.** Total transmission (normalized to the maximum of the curve) as a function of the cluster size $k_{max}$ with the introduction of layers of a third material with refractive index $n_3$ randomly distributed in the photonic structure. The percentage of defects in the legend corresponds to the percentage of $n_3$ layers in the structures (e.g. 2% relates to 16 $n_3$ layers over the 800 layers composing the structures). In this study $n_1 = 1.4$; $n_2 = 2.0$; $n_3 = 2.2$.

In Figure 20 we report this study for different numbers of defects in the structure. We observed that only a heavy introduction of defects (≥ 10%) affects significantly the trend. Increasing the number of defects, the magnitude of transmission valley decreases and the oscillations are strongly affected.

*5.1.4. Brief discussion on Shannon index and distributions of clusters*
Our results show how the disorder of the photonic structure affects the total light transmission. Focusing on Figure 13, Figure 15 and Figure 17, i.e. the behaviour of the normalized total transmission as a function of the Shannon index $H'$, the power law $a$, and the maximum cluster size in a uniform distribution $k_{max}$, respectively, we could try to draw a picture, taking into account the differences of the three studies.

Thus, the degree of disorder of the photonic structure, measured by three different methods, determines the total light transmission with a common pattern. The total transmission decreases as a function of the Shannon index, decreases as a function of the power law exponent, increases, after a valley, as a function of the maximum cluster size in a uniform distribution. In information theory the more homogeneus structure, where the elements are distributed on the entire surface, are the structure with higher disorder. Considering this definition of disorder in our photonic structures, the higher is the Shannon index, the power law exponent and the $k_{max}$ of the uniform distribution, the higher is the disorder.

For this reason, we can say that in our experiments we observe a common pattern where the more aggregated (less homogeneus) is the structure, the higher is the light transmitted.

The Shannon index is widely used to quantify, for example, the spatial distribution of species in ecology. It describes the homogeneity of the number or species along samples area, i.e. when the number of species is the same in every sample area, the Shannon index has the highest value [226]. Focusing on the distribution of a certain kind of objects in a structure, the Shannon index can be viewed as a general measure of the homogeneity of that structure. In our study the Shannon index indicates the homogeneity of the photonic crystal focusing on the layer distribution, i.e. when the layers are equally distributed on the spatial structure of the photonic crystal, the Shannon index is the maximum.

The Shannon index is a measure of the homogeneity of the structure, thus less homogeneous structures transmit more light. The fact that the total transmission increases with $k_{max}$ is in good agreement with the latter, since a higher $k_{max}$ (more large clusters) leads to a higher non homogeneity. There is an agreement also with the fact that the total transmission decreases with $a$, because a higher $a$ strongly decreases the probability to have large clusters (that leads to a higher non homogeneity).

It is noteworthy that, for the first values of $k_{max}$, the total transmission decreases up to a valley (Figure 17), as mentioned above. This behaviour could be ascribed to the fact in this range of $k_{max}$ there is a trade-off between the maximum cluster size, that decreases the total transmission, and the non homogeneity (disorder) of the medium, that increases the total transmission. In particular, from $k_{max} = 1$ to $k_{max} = 4$ we observed a decrease in the total transmission (Figure 17, 18, and 19), and this can be ascribed to the fact that the light is more efficiently reflected (i.e. less transmitted) by thicker high refractive layer clusters; when $k_{max} > 4$, the higher transmittance due to the homogeneity of the structure becomes instead predominant.

All these studies indicate that the presence of large refractive layers clusters reduces the transmission of light in photonic crystals. These results show something new with respect to the literature, in which the light transmission is studied paying attention to the sample of the medium [45,49]. Not only the sample length plays a major role, but also the aggregation of the high refractive index layers in media (i.e. the length of each high refractive index region in the 1D structure).

Our simulations demonstrate that is possible to investigate the total light transmission of a photonic medium correlating his degree of disorder with a statistical ensemble. This can be useful when deterministic and correct computation of the total light transmission are not feasible.

*5.2. Fabrication techniques*

The fabrication of controlled disordered photonic crystal is a technological and scientific challenging task. To achieve the correct thicknesses in disordered 1D photonic structures, in fact, a strictly monitor of each layer is mandatory but, on the other hand, the fabrication technique have to allow the realization of homogeneous films with thicknesses that could reach some hundred nm if application in visible/NIR regions is required using dielectric materials.

Rf-sputtering is a possible choice as fabrication technique for such structures thank to the capability to obtain high reproducibility reproducibility of the single layers and to monitor the deposition rates during the process. Moreover, the ability of this technique to process various materials like dielectric systems make it a great candidate for the fabrication of glass based disordered 1D photonic structures [62].

*5.3. Applications*

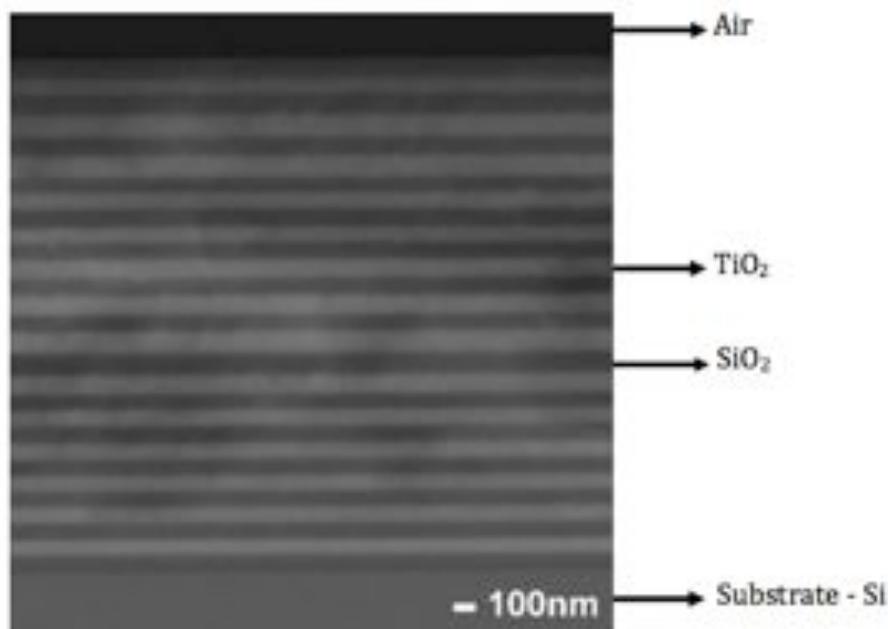

**Figure 21.** SEM micrograph of a photonic structure made by alternating $SiO_2$ and $TiO_2$ layers with random thickness.

Figure 21 shows the SEM micrograph of the microcavity composed by 14 pairs of $TiO_2/SiO_2$ layers. To realize the disordered photonic structure, we have alternated layers of $SiO_2$ and $TiO_2$, with a thickness of $(80 + n)$ nm, where n is a random integer $0 < n < 40$, by multi target rf sputtering technique using silicon and silica substrates. In this way we obtain a random sequence of thicknesses, between 80 and 120 nm. To monitor the thickness of the layers during the deposition, two quartz microbalances Inficom thickness monitor model SQM 160, faced on the two targets have been employed. For the deposition thickness monitor was calibrated for the two kinds of materials by a long deposition process (24h of deposition) and by directly measuring the thickness of the deposited layer by an m-line apparatus and SEM imaging. The final resolution on the effective thickness obtained by this quartz microbalance is about 1 Å.

As shown in Figure 22 such structures have the advantage to exhibit a broad transmission band and lower transmittance with respect to the corresponding periodic photonic crystal, opening the way to the fabrication of broad band filters. In the case of the example shown in figure 21, an average transmittance value of 0.7 was obtained for the 300 – 1200 nm transmission spectrum reported in figure 4 [62].

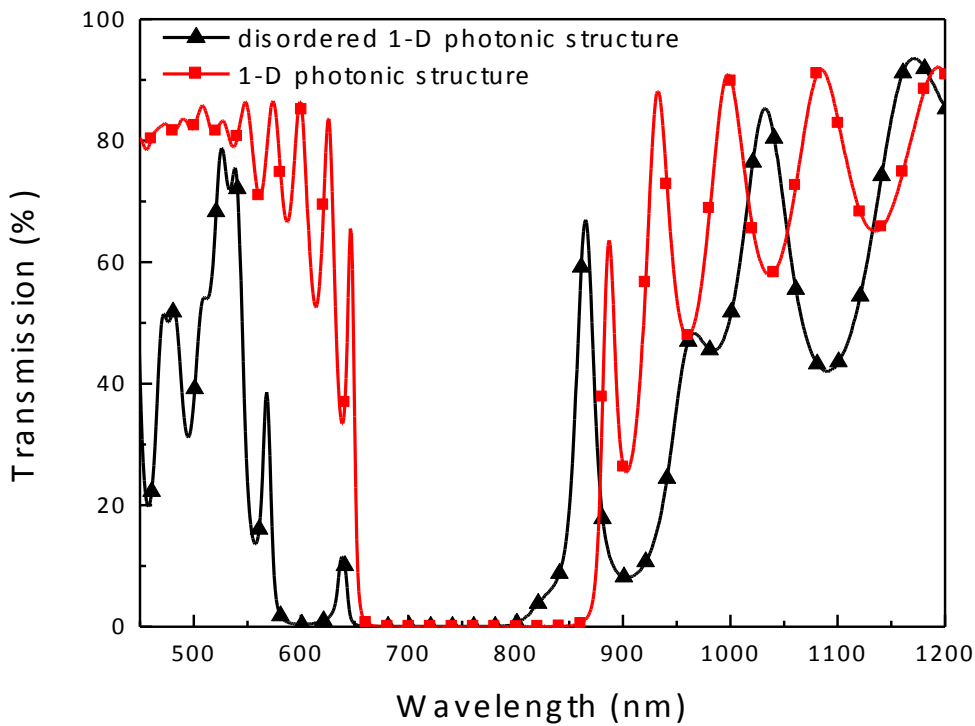

**Figure 22.** Transmission spectra obtained from a (■) 1D photonic crystal and a (▲) disordered 1D photonic structure (b).

Figure 23 shows the direct comparison of the reflectance properties of one dimensional photonic crystals (a) and disordered 1D photonic structures (b). The only difference is the randomness in thickness. This appealing behaviour is due to the interference between waves traveling in regions with different optical paths, determined by the disordered distribution of stacked layer thicknesses.

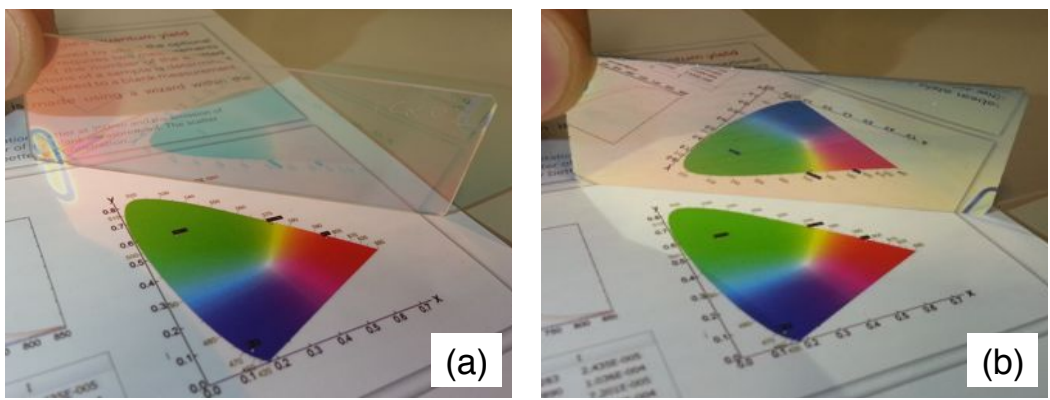

**Figure 23.** Transmission spectra obtained from (a) a 1D photonic crystal and (b) a disordered 1D photonic structure. Adapted from Ref. [62].

Random lasing is an other application that employs disorder photonic structures [31,227]. In the one-dimensional case, Milner and Genack demonstrated random lasing in a stack of partially reflecting glass slides, air gaps between the slides, and amplifying regions consisting either of Rhodamine 6G solutions between the slides or a single chromophore-doped plastic sheet [32].

Finally, a non-periodic photonic structure have been integrated in a polymer solar cell to enhance the light trapping and preserve the cell transparency for a possible incorporation of such cells into buildings [228].

## 6. Conclusions

In this review article we discussed the optical properties of 1D periodic, aperiodic, and disordered structures. We have described some theoretical methods to study the structures, also discussing the refractive indexes of the most used materials. We then discussed the optical response, the fabrication methods, and some of the most significant applications of the 1D structures, as sensing, light emission enhancement and light trapping for photovoltaic devices. Focussing on the optical response of 1D disordered photonic structures, we discussed different types of disorder in 1D photonic structures. First of all, we demonstrated that the homogeneity of a photonic structure can be quantified by the Shannon index and it is possible correlated the normalized total transmission with the Shannon index. Second, by aggregating the high refractive index layers in clusters with the size following a specific distribution, either power law or uniform, we could relate the normalized total transmission to: i) the exponent $a$ of the power law; ii) the maximum cluster size of the uniform distribution. Such studies open the way to a better understanding of the light transmission through disordered photonic structures, showing that the light transmission follows some specific behaviours if the disorder has some peculiar characteristics (in terms of arrangement in the unit cells or cluster size). These studies can be useful for the employment of such disordered media for photovoltaics and photodetection. Finally, we discussed some applications with 1D disordered photonic structures.

## Appendix
A)

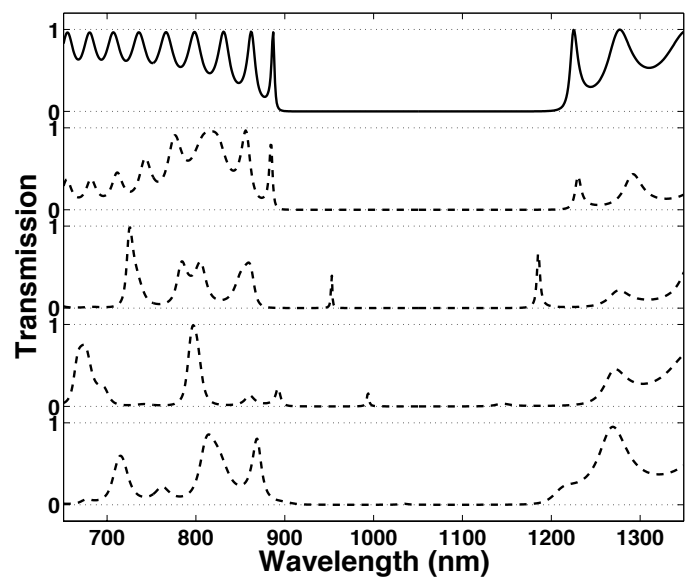

**Figure A1.** Transmission spectrum of a periodic photonic crystal, indeed with $H'$ = 1, of 16 unit cells $TiO_2$-$SiO_2$-$SiO_2$-$SiO_2$ (solid line) and transmission spectra of disordered structures with $H'$ = 1 (dashed lines).

B) Script for the transfer matrix method to calculate the transmission spectrum of a 8-bilayer 1D photonic crystal (for MATLAB). The materials employed have relative permeabilities $\mu_j$ = 1. In principle, the script can be written in any numerical computing software language.

```
%Transfer Matrix Method – air-multilayer-glass
clear;clc;
n0=1; %refractive index of air
ns=1.46; %refractive index of glass substrate
n1=2;d1=100; %refractive index and thickness (nm) layer 1
n2=1.6;d2=100; %refractive index and thickness (nm) layer 2
M1=[0, 0; 0, 0];M2=[0, 0; 0, 0];ii=0;
for i=300:1:1300 %wavelength range and step (in nm)
  ii=ii+1;
  l(ii)=i; %wavelength
  S1(ii)=cos(2*pi*n1*d1/l(ii));
  P1(ii)=-(1i*sin(2*pi*n1*d1/l(ii)))/n1;
  Q1(ii)=-1i*n1*sin(2*pi*n1*d1/l(ii));
  R1(ii)=cos(2*pi*n1*d1/l(ii));
  M1=[S1(ii), P1(ii); Q1(ii), R1(ii)];%matrix 1 for layer 1
  S2(ii)=cos(2*pi*n2*d2/l(ii));
  P2(ii)=-(1i*sin(2*pi*n2*d2/l(ii)))/n2;
  Q2(ii)=-1i*n2*sin(2*pi*n2*d2/l(ii));
  R2(ii)=cos(2*pi*n2*d2/l(ii));
  M2=[S2(ii), P2(ii); Q2(ii), R2(ii)];%matrix 2 for layer 2
  M=(M1*M2)^8;%matrix product with number of bilayers
  t=2*ns/(((M(1,1)+M(1,2)*n0)*ns)+(M(2,1)+M(2,2)*n0));
  T(ii)=(n0/ns)*(abs(t)^2);%transmission
end
plot(l,T,'k');
```

**Acknowledgements**
This project has received funding from the European Union's Horizon 2020 research and innovation programme (MOPTOPus) under the Marie Skłodowska-Curie grant agreement No. [705444], as well as (SONAR) grant agreement no. [734690].